# Exponential Competence of Computer Science and Software Engineering Undergraduate Students

Orit Hazzan
*Faculty of Education in Science and Technology*
*Technion – Israel Institute of Technolohu*
Haifa, Israel
oritha@ed.technion.ac.il

*Abstract*—We live in exceptional times in which the entire world is witnessing the exponential spread of a pandemic, which requires to adopt new habits of mind and behaviors. In this paper, I introduce the term exponential competence, which encompasses these cognitive and social skills, and describe a course for computer science and software engineering students in which emphasis is placed on exponential competence. I argue that exponential competence is especially important for computer science and software engineering students, since many of them will, most likely, be required to deal with exponential phenomena in their future professional development.

## I. Introduction

Over the last several months, the term *exponential growth* has practically become a household concept throughout the world. Without going into details regarding the general public's understanding of the concept, this paper aims to highlight the importance of computer science and software engineering undergraduate students understanding exponential growth, as well as associated terms. Specifically, in this paper, I introduce the term *exponential competence* and describe a course on soft skills for computer science and software engineering students in which emphasis is placed on exponential competence.

Competence, in general, is the ability to perform something proficiently, taking a comprehensive perspective, and demonstrating deep understanding of its context; *exponential competence*, specifically, means being able to adapt to fast and unpredictable changes in the environment and dealing well with exponential phenomena, such as the Corona pandemic, the growth of exponential organization[1] [3], the speed of innovation adoption, and the growth of job opportunities for data scientists.

As all competences, exponential competence encompasses knowledge, skills, attitudes and behaviors, many of which are considered to be "soft" [1, 4]. *Soft skills* is a term that refers to skills that are required when interacting with, working with, and managing people, and hence includes cognitive skills, such as problem solving, as well as social skills, such as communication. Soft skills are considered essential for computer science and software engineering students for their future professional development.

Exponential competence is especially important for computer science and software engineering students, since many of them will, most likely, be required to deal with exponential phenomena in the future, either as entrepreneurs or as employees in organizations whose growth is exponential or, like the general public, as consumers of products produced by such organizations.

Competences are made up of four components: knowledge, skills, attitudes, and behaviors. The knowledge component of exponential competence relates to computer science and software engineering topics such as algorithms, computability, complexity, software development processes, testing, Internet of Things (IoT), cloud computing, and data science. This component is not discussed in this paper, which focuses rather on the skills, attitudes, and behavior components of exponential competence, describing how they are addressed in a course on soft skills that was offered to undergraduate computer science and software engineering students at two leading universities in Israel.

The first course was taught F2F at the Technion – Israel Institute of Technology in the Spring 2018 semester. It included a hackathon designed to foster the students' exponential competence, in which the students were asked to prototype a product for people with disabilities, that has the potential of being adopted via exponential growth. The second course was taught during the first Corona semester (2020 Spring semester) at the Hebrew University of Jerusalem's School of Engineering and Computer Science (as part of my sabbatical). Although this course is being taught online due to the social distancing regulations, exponential competence was receiving special attention in the course due to these exceptional times in which the students, together with the entire world, are witnessing the exponential spread of the pandemic.

## II. Course description

Both courses were elective and were based largely on active learning in teams, either F2F (at the Technion) or in breakout rooms (at the Hebrew University of Jerusalem). Most of the students in the two courses were in their 3rd or 4th year of studies. The Technion course was taught with the assistance of two TAs; the HUJI course is being taught by me alone. Further details on the two courses and their student populations are presented below in Table I.

---

[1] Companies, such as, Amazon, Uber, Spotify and Netflix, which grow at an above-average rate – up to ten times faster than comparable companies in the industry – but can make do with considerably fewer resources thanks to new forms of organization and the use of new, especially digital, technologies.

## A. Course description - Technion

The course met during the 2018 Spring semester for three weekly hours. 46 students attended the course: 36 CS major students, 9 CS education major students, and 1 electrical engineering student. 36 students were males and 12 – females (as is the gender distribution in the Technion's Faculty of CS). Most of the students were juniors or seniors. The soft skills course was elective for CS major students and the engineering students, and mandatory for students who studied CS education program.

The course was divided into three parts:

Lessons 1-3 – Pre-hackathon: In this part of the course, focus was placed on common soft skills such as teamwork, presentations, and giving and receiving feedback.

Lessons 4-9 – The hackathon: This part included the hackathon as well as the preparation stage and several reflective assignments, in which students analyzed the initiatives they developed in the hackathon from an exponential perspective.

During the hackathon, students were introduced to various challenges faced by people with physical, cognitive, and learning disabilities, attention deficit problems, and so on. Since the proposed challenges are faced by real communities, the hackathon also enabled students to practice finding solutions to specified needs, and then planning and implementing them so as to fulfil those specific needs, all of which are activities included in exponential competence.

The focus on exponential competence during the hackathon enabled students to enhance various related skills, attitudes, and behaviors in a stressful and competitive environment, while dealing with social challenges. Unlike short exercises facilitated during the regular lessons, special events, such as hackathons, further highlight the importance of skills, such as listening, communication, emotional intelligence, and teamwork, of attitudes, such as flexibility, and of behaviors, such as working under pressure, all of which fall under the umbrella of exponential competence.

Lessons 10-13 – Post-hackathon: In this final part of the course, we addressed additional soft skills, further reflecting on the mutual relationships between soft skills and exponential competency.

## B. Course description – Hebrew Univerity of Jerusalem

This course, which was taught online during the 2020 Spring (Corona) semester, as part of my sabbatical at the School of Engineering and Computer Science at the Hebrew University of Jerusalem, was an opportunity to leverage the discussion on exponential competency due to the wide spread use of the term exponential growth in the context of the pandemic. This enabled to raise the level of abstraction of the discussion about exponential phenomena beyond the context of computer science and software engineering and to explore skills, attitudes, and behaviors required in order to deal with exponential phenomena. Associated terms, such as internationality, mobility, and cultural intelligence, were also being addressed in the course. Zoom features, such as the chat and breakout rooms, were facilitated regularly in all lessons, not only fostering teamwork, but also exposing students and training them to work remotely in either globally or locally distributed teams.

The course is divided into three parts:

- Lessons 1-2: Introduction and discussion about exponential phenomena in general, and exponential organizations in particular.

- Lessons 3-8: In each of these six lessons, three pairs of students each present a soft skill according to their choice. Following each presentation, which lasts ~15 minutes, the student pairs receive feedback from their classmates (5 minutes). This is done in a forum in which the students were asked to address a question posed by the presenters. The presenters were asked to summarize these feedbacks and to briefly present this summary in the following lesson (3 minutes). In addition, after the three student presentations, additional soft skills relevant for computer scientist and software engineers are presented in each lesson, either by a guest lecturer or by me, including a discussion of their global context and their exponential context, if relevant.

- Lesson 9: Summary – warp up, main messages and future options for professional development.

35 students attended the course: 26 CS major students, 7 Internet and Society students, and 2 dual major students. 19 students were males and 16 – females. Most of the students were juniors. The soft skills course was elective for all.

Table II presents examples of the tasks given to the students in the two courses to further foster their exponential competence.

TABLE I. MAIN CHARACTERISTICS OF THE COURSE POPULATIONS

| University | Semester | Number of weekly hours | Teaching method | Student population – Number and programs | Special characteristics |
|---|---|---|---|---|---|
| Technion – Israel Institute of Technology | Spring 2018 | 3 for 13 weeks | F2F | 46:<br>- 36 Computer Science students<br>- 9 prospective Computer Science teachers<br>- 1 Electrical Engineering student | Hackathon in mid- semester in which the concept of exponential competence was introduced. |
| Hebrew University of Jerusalem | Spring 2020 | 3 for 9 weeks (instead of 2 for 13 weeks) | Online | 35:<br>- 26 Computer Science students<br>- 7 Internet and Society students<br>- 2 dual major students | Exponential competence was introduced from the onset of the course and was highlighted throughout the course on relevant occasions. |

TABLE II. EXAMPLES OF COURSE TASKS GIVEN TO FOSTER EXPONENTIAL COMPETENCY

| Technion F2F course tasks | Hebrew University of Jerusalem online course tasks |
|---|---|
| *Team reflection on the hackathon*<br>As a team, complete a questionnaire in which you analyze, from a soft skills' perspective, the initiative the team developed in the hackathon as an exponential organization.<br>Each team shall submit one reflection.<br>Following are some representative questions:<br>• Indicate 10 soft skills that your team implemented in the hackathon. For each skill, describe how you implemented it and for what purpose.<br>• Indicate 10 soft skills that your team failed to implement. For each skill, describe how it would have contributed to the development of your initiative.<br>• Indicate 10 soft skills that employees of exponential organizations should express. | *In an introductory questionnaire (before the onset of the semester)*<br>a. In your opinion, what soft skills are especially important for coping with the current pandemic?<br><br>*Teamwork in breakout rooms – Lesson 2*<br>b. We will work in 10 breakout rooms. Each room will be allocated one characteristic of exponential organizations.<br>  o Select an exponential organization and analyze it according to the characteristics allocated to your room.<br>     Examples of exponential companies can be found here: http://top100.exponentialorgs.com/<br>  o Indicate at least 5 soft skills required for the implementation of this characteristic.<br>c. Indicate 10 soft skills that employees of exponential organizations should express.<br><br>The skills mentioned by the students in their answers to sections a. (51 skills) and c. (76 skills) were collected and presented back to them in two lists. The students worked on the following task:<br>d. Select 5 soft skills that appear on both lists and explain its importance and expression in the context of exponential phenomena. |
| *Questions presented to the students in both courses* | |
| - List 5 advantages of working in an exponential organization.<br>- List 5 disadvantages of working in an exponential organization.<br>- List 3 cognitive challenges that employees in exponential organizations likely face.<br>- List 3 social challenges that employees in exponential organizations likely face.<br>- On a scale of 1 (not at all) to 4 (very much), to what degree would *you* like to work in an exponential organization? | |

## III. ANALYSIS OF STUDENTS' ANSWERS FROM THE EXPONENTIAL COMPETENCE PERSPETIVE

The richness of the collection of soft skills mentioned by the students in their answers to the above questions enables to sort them in different ways, for instance, according to

- the *MERge* model for professional development, which encompasses three meta-professions, i.e. management, education, and research, beyond the profession of the practitioner [2].
- different perspectives:
  ✓ cognitive – understanding what exponential growth means and solving relevant problems;
  ✓ social – awareness of the potential, as well as risks, of exponential social influences;
  ✓ organizational – evaluation the impact of a specific product as linear or exponential, and awareness of the impact of the organization characteristics on its growth;
  ✓ technological – resources and techniques needed to implement exponential technologies; and more.

However, as I discuss in this paper, the students' answers were sorted according to the components of competence. Table III presents the capabilities students from the two courses mentioned in their responses to the above questions, divided into skills, attitudes, and behaviors and on three levels – individual, team and organization. Although most skills were mentioned multiple times in each of the two courses, they are each presented here only once.

As can be seen:

- Although the students were asked to list soft skills in response to the above questions, the richness of their answers enables to sort them into the components of competence: skills (the ability to do something well), attitudes (ways of thinking or feeling about someone or something), and behaviors (way in which people act).

- The boundaries between the three components are not always clear-cut, but the richness of the exponential competence is expressed clearly. Furthermore, this richness enables to categorize the skills, attitudes and behaviors into three levels – the individual, the team, and the organization.

- Although the two courses were taught and learned via different media (the first used F2F interaction and the second was conducted completely online), the attention given in both cases to exponential phenomena focused the students' attention on the skills, attitudes and behavior components of the competence.

- Table III does not include the knowledge component of competence since students in these two courses tend to distinguish between content they acquired in other courses (knowledge in terms of competence) and content studied in the said courses (skills, attitudes and behavior in terms of competence).

TABLE III. SKILLS STUDENTS MENTIONED WHEN DEALING WITH EXPONENTIAL PHENOMENA SORTED ACCORDING TO THE COMPONENTS OF COMPETENCE AND BY ORGANIZATION LEVEL

|  | Skills | Attitudes | Behaviors |
|---|---|---|---|
| **Individual** | <ul><li>Ability to analyze data effectively, decide whether or not the resource is reliable, ability to absorb and process data, draw conclusions from information, information management, knowledge management, ability to distinguish between important and unimportant or between true and false</li><li>Fast learning, learning from mistakes, quick thinking, problem solving</li><li>Entrepreneurship, aesthetic sense, ccreativity, identifying opportunities</li><li>Charisma, self-confidence, self-awareness, persuasion skills, self-propelled ability</li><li>Time management</li><li>Personal responsibility, critical thinking, self-criticism</li><li>Decision making</li><li>Emotional intelligence, empathy, aattention, llistening, ppatience</li><li>Persistence, self-discipline in studies, working from home</li></ul> | <ul><li>Multidisciplinary</li><li>Ability to deal with situations of uncertainty, adaptation to change, agility</li><li>Openness to criticism</li><li>Motivation for continuous learning and professional development</li><li>Flexible thinking</li><li>Positive attitude</li><li>Cool</li><li>Rigor</li><li>Optimism</li><li>Modesty</li><li>Tolerance</li><li>Professionalism</li><li>Independence</li></ul> | <ul><li>Presentation skills</li><li>Writing skills</li><li>Negotiation</li><li>Working under pressure</li><li>Order</li><li>Give and receive feedback</li><li>Enrolling people to tasks</li><li>Setting goals</li><li>Taking risks</li><li>Transparency</li><li>Code sharing</li><li>Accepting authority</li><li>Ability to deal with failures</li><li>Attention to details</li><li>Defining clear tasks</li><li>Use of technological tools</li><li>Negotiation management</li><li>Effectivity in stimuli-rich environments</li><li>Task prioritization</li></ul> |
| **Team** | <ul><li>Teamwork, persuasion abilities</li><li>Ability to coordinate teams</li></ul> |  | <ul><li>Ability to create trust among users</li><li>Collaboration with others, team play</li><li>Interpersonal communication</li><li>Good personal relationships</li><li>Ability to work in diverse teams</li></ul> |
| **Organization** | <ul><li>Manage and integrate multiple entities, responsibility for various aspects</li><li>Ability to manage and decentralize tasks, employee and task management, marketing ability</li><li>Business vision</li><li>Global-scale synchronization</li><li>Creating community</li><li>High social responsibility</li></ul> | <ul><li>Commitment and involvement</li><li>Environmental sensitivity</li><li>Work ethics</li></ul> | <ul><li>Change management, leadership</li><li>Remote management, crisis management</li><li>Strategic management</li><li>Budget planning and budget compliance, resource management</li><li>Ability to understand customer needs</li><li>Decentralization of powers</li></ul> |

## IV. SUMMARY

In the summary, I discuss the exponential competence from the perspective of Israel, which is known as the Startup Nation [5].

Many multinational companies have established local sites in Israel [2], and many of them, like Google, Facebook and Amazon, are exponential. Other exponential companies currently operating in Israel were founded by Israeli entrepreneurs, e.g., Waze (sold to Google) and Mobileye (sold to Intel).

Although the exponential company Zoom Video Communications is not active in Israel, its case is especially interesting, not only because Oded Gal, Zoom's Chief Product Officer, is a Technion graduate, but also because we are discussing exponential competence, and Zoom has demonstrated exponential growth during the Corona pandemic. According to Oded Gal, while in December 2019 Zoom was being used by 10 million meeting participants daily, at the end of April 2020, this number had grown to 300 million a day.

Graduates of Israel's leading universities are largely involved in this *Startup Nation* scene in general, and in its exponential facets in particular. This is true also for the students who studied the courses described in this paper. When asked to rate, on a scale of 1 (not at all) to 4 (very much), the degree to which they would like to work in an exponential organization, 80%-90% of both classes expressed interest (rated their interest 3 and 4). It is, therefore, important to enable students to acquire exponential competence, which as we saw, encompasses many skills, attitudes, and behaviors needed in order to cope with exponential phenomena.

REFERENCES

[1] L. Carter, "Ideas for Adding Soft Skills Education to Service Learning and Capstone Courses for Computer Science Students". In Proceedings of the 42nd ACM technical symposium on Computer science education (SIGCSE '11). ACM, New York, NY, USA, 2011, pp. 517-522.

[2] O. Hazzan and R. Lis-Hacohen, R, The MERge Model for Business Development: The Amalgamation of Management, Education and

---

[2] See https://en.wikipedia.org/wiki/List_of_multinational_companies_with_research_and_development_centres_in_Israel

Research, SpringerBriefs in Business, 2011. http://www.springer.com/us/book/9783319302249

[3] S. Ismail, M.S. Malone and Y. van Geest, Exponential Organizations: Why new organizations are ten times better, faster, and cheaper than yours (and what to do about it), Diversion Books, 2014.

[4] G. Matturro, F. Raschetti and C. Fontán,. Soft skills in software development teams: a survey of the points of view of team leaders and team members. InProceedings of the Eighth International Workshop on Cooperative and Human Aspects of Software Engineering (CHASE '15). IEEE Press, Piscataway, NJ, USA, 2015, pp. 101-104.

[5] D. Senor and S. Singer, S, Start-up Nation: The Story of Israel's Economic Miracle, Twelve, 2008.